\documentclass[aip,rsi, amsmath,amssymb,reprint,]{revtex4-1}

\usepackage{chemformula} 
\usepackage{url}
\usepackage[english]{babel}

\usepackage{chemist}
\usepackage{blindtext}
\usepackage{lipsum}
\usepackage{graphicx}
\usepackage{color}

\usepackage[normalem]{ulem}
\usepackage{soul}

\begin{document}

\title{Solvent-mediated interactions between nanostructures:\\
   from water to Lennard-Jones liquid}

\author{Julien Lam}
\email{julien.lam@ulb.ac.be}
\author{James F. Lutsko}

\affiliation{Center for Nonlinear Phenomena and Complex Systems, Code Postal 231, Universite Libre de Bruxelles, Boulevard du Triomphe, 1050 Brussels, Belgium}

\begin{abstract}
Solvent-mediated interactions emerge from complex mechanisms that depend on the solute structure, its wetting properties and the nature of the liquid. While numerous studies have focused on the two first influences, here, we compare results from water and Lennard-Jones liquid in order to reveal to what extent solvent-mediated interactions are universal with respect to the nature of the liquid. Besides the influence of the liquid, results were obtained with classical density functional theory and brute-force molecular dynamics simulations which allows us to contrast these two numerical techniques. 
\end{abstract}

\maketitle

\section{Introduction}

Interactions between two solids are usually well characterized by their intrinsic physical and chemical properties. However, in the presence of a liquid solvent, additional interactions emerge and can become dominant when the solids are also electrically neutral. This so-called solvent-mediated interaction is involved in various phenomena including self-assembly\cite{Leikin1994,Chandler2005,Kowalik2017,Maibaum2004}, ligand unbinding\cite{Tiwary2015a,Tiwary2015b,Jamadagni2009} and protein folding\cite{Camilloni2016,Spolar1989}.

Numerous studies have examined solvent-mediated forces in terms of range, strength and sign using both numerical\cite{Li2005,Patel2010,Patel2011,Huang2001,Banerjee2015,Kanduc2015,Kanduc2016,Chacko2017,Lam2017,Stock2017} and experimental techniques\cite{Pashley1985Sep,Hato1996Jan,Meyer2006,Dishon2009,Mastropietro2012Mar,Azadi2015Feb,Schlesinger2017Mar,Ishida2018Feb,Marra1985Aug,Israelachvili1983Nov,Chen2016Jun}.  In particular, it was found that the sign of solvent-mediated forces is controlled by the equilibrium contact angle\cite{Kanduc2016,Kanduc2015,Chacko2017,Lam2017,Hato1996Jan}. On the one hand, when the solids are solvophilic, solvent molecules are attached to the solid surface. Bringing the two solutes together leads to a perturbation and the removal of this favorable structure causes a strong repulsive hydration pressure\cite{Marra1985Aug,Israelachvili1983Nov,Chen2016Jun}. On the other hand, when the solids are solvophobic, solvent molecules which are located between the two solids are expelled and a more stable vapor cavity emerges thus reducing the overall free energy.  This so-called capillary evaporation has been the subject of numerous works\cite{Vaikuntanathan2016Apr,Bolhuis2000Nov,Altabet2017Mar,Remsing2015Jul,Berard1993May,Evans2017Jul} and leads to a solvophobic attraction. Along with the wetting properties of the solid, the role of its geometrical structure was also covered in several studies\cite{Qin2003Nov,Jabes2016Aug,Qin2006Feb}. For example, Jabes et al. recently demonstrated that using the same solid composition and size, qualitatively different solvent-mediated forces can be obtained only by changing the solid shape between fullerenes, nanotubes and graphene-like structures\cite{Jabes2016Aug}. 
Altogether, this work is integrated to a general mean-field theory of hydrophobicity developed by Lum, Chandler and Weeks \cite{Lum1999Jun} and further refined in more recent publications\cite{Varilly2011Feb,Remsing2013Dec,Vaikuntanathan2016Apr}. 

When modeling liquids using numerical simulations, two approaches are commonly employed. On the one hand, because water plays a crucial role in most applications and especially in biological systems, atomistic models for water molecules have been developed in order to mimic its thermodynamic properties and some special features including strong hydrogen bonding and ice polymorphism. On the other hand, generic model systems such as hard spheres\cite{asakura1954,dickman1997,roth1999,xiao2005,nygaard2016a,nygaard2016b,mishima2013a,
mishima2013b,hara2016} and Lennard-Jones potential\,\cite{stewart2014,Chacko2017,maciolek2004} are also often used for modeling fluids. Once the model is chosen, solvent-mediated forces can be computed using various types of numerical methods. Monte-Carlo and molecular dynamics simulations are widely employed especially for water modeling while classical density functional theory (DFT) in which molecules are treated as a density field can grant access directly to liquid density profiles and the corresponding free energy\cite{Lutsko2010Jan,Lam2017}. DFT is less numerically expensive and avoids using free energy calculation techniques such as thermodynamic integration, transition path sampling and umbrella sampling. However, DFT for water is not as highly developed as for simple fluids\cite{Jeanmairet2013Feb,Jeanmairet2013Oct,Hughes2013Jan}.

While numerous authors have suggested the ability of the Lennard-Jones liquid to reproduce behaviors similar to water regarding solvent-mediated effects\cite{Ashbaugh2013Aug,Evans2015Apr,Evans2015Jul}, there is no detailed comparison of solvent-mediated forces obtained with atomistic simulations of extended simple point charge model water and with DFT calculations of Lennard-Jones particles (LJ). In this work, we make a direct comparison of solvent-mediated forces obtained from molecular dynamics simulations of water and DFT calculations of LJ using a very generic system made of two nanometric crystalline slabs immersed in a liquid. Free energy is computed as a function of the interslab distance and we study thoroughly the influence of wettability and of the slab geometrical structure. Our work identifies differences and similarities between atomistic simulations of water and DFT calculations of LJ. Moreover, our results also contribute to the overall understanding of solvent-mediated forces and discuss more generally to what extent molecular properties of water make it special in comparison to simple fluid models.

\section{Methods}

\subsection{Studied system}

Our calculations make use of two types of molecules: slab molecules and liquid molecules. The slabs are composed of rigid arrangements of solid molecules while the liquid is treated dynamically. We held constant temperature and density of the liquid while varying the solid properties. The slabs are made of three square layers of $40+41+40=121$\,atoms which are kept fixed in a face centered cubic structure with the (100) face exposed and with the lattice spacing, $a$. The interaction between slab and liquid particles is modeled via a Lennard-Jones potential parametrized by its length scale, $\sigma_{wall}$ and well depth, $\epsilon_{wall}$. When varying $a$, $\sigma_{wall}$ is also modified using: $\sigma_{wall}=\frac{a}{a_0} \sigma_{liq}$ where $a_0$ is the zero-temperature FCC equilibrium lattice spacing equal to $1.5424\, \sigma_{liq}$\,\cite{Ashcroft}. With such a model, wetting properties as defined by the contact angle are driven by the ratio between the liquid/liquid and liquid/solid attractions. In practice, we varied $\epsilon_{wall}$ while keeping the liquid properties constant and measured the corresponding contact angle, $\theta$. We note that additional complexities that also influence the surface solvophobicity including functionalization and polarity effects can not be captured with our present model\cite{Jabes2016Aug}.

Finally, results are shown in physical units. For water, energy is displayed in kcal/mol and distances in angstroms. When computing $a$, we used $\sigma_{liq}=2.75\,$\AA\,as it is the approximate size of a water molecule. For LJ, the potential parameters are denoted $\epsilon_{LJ}$ and $\sigma_{LJ}$. We worked at a temperature of $k_BT=0.8\epsilon_{LJ}$ and we chose: $k_BT=0.593$\,kcal/mol at $300$\,K in order to rescale $\epsilon_{LJ}$ to real units.\,Concerning the distances, we imposed $\sigma_{LJ}=2.75\,$\AA\,as well.

\subsection{Molecular dynamics simulation of water}

The extended simple point charge model\,(SPC/E) is used for water\cite{Berendsen1987Nov}. Bonds in water molecules are constrained using the SHAKE algorithm and long-range Coulombic interactions are computed with the Particle-Particle-Particle-Mesh solver with a precision tolerance equal to $10^{-4}$ and a real space cutoff equal to $9.8$\,\AA. At the initialization step, water molecules are arranged on a simple cubic lattice structure with a lattice spacing equal to $3.1$\,\AA. Solids are modeled with rigid molecules made of three face-centered cubic layers. Solid molecules and oxygen atoms of water interact via a truncated and shifted Lennard-Jones (LJ) potential with a cutoff equal to $9.8$\,\AA. The LAMMPS package\citep{plimpton1995} is used for all  the simulations.  From there, two types of calculations are performed: (i) Droplet equilibration to measure the wetting properties of the solid and (ii) Solvent-mediated forces between two slabs.

\subsubsection{Droplet equilibration and contact angle}

\begin{figure}[h!]
\includegraphics[width=8.6cm]{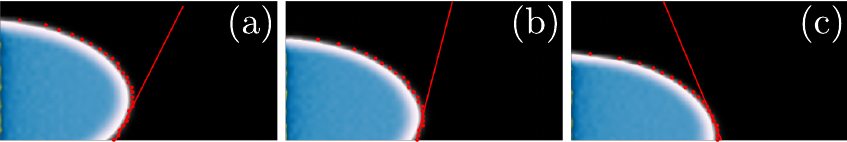}
\caption{Water density profiles for an equilibrated water droplet on top of a wall with $\epsilon_{wall}$ equal to $0.1$\,kcal/mol (a), $0.2$\,kcal/mol (b) and $0.3$\,kcal/mol (c). Red dots represent the liquid-gas interface which is used to compute the contact angle and red lines are the associated linear fit. The image size is equal to $30$\,\AA$\times 60$\,\AA.}
\label{DensityDroplet_DFT}
\end{figure}

A hemisphere of water with radius equal to $50$\,\AA\,is initially deposited onto the solid surface. On top of the spherical cap, a cage made of fixed atoms is also placed to help the droplet equilibration and prevent it from leaving the solid surface at the initialization stage. These atoms only interact with oxygen atoms via a LJ potential ($\epsilon_{cage}=1$\,kcal/mol and $\sigma_{cage}=\sigma_{O-O}=3.166$\AA). The entire simulation box measures $300$\,\AA$\times 300$\,\AA$\times 200$\,\AA\,which is large enough to avoid the influence of periodic images.  For the equilibration protocol, the timestep is set equal to $0.5$\,fs. NVE simulations are performed during $5$\,ps then NVT simulations are performed during another $5$\,ps at $300$\,K. From there, cage atoms are removed to allow for droplet shape relaxation and the timestep is changed to $2$\,fs. After an  equilibration run during $500$\,ps, snapshots are taken every $ 2.5$\,ps during $500$\,ps. Density profiles of oxygen atoms are averaged through time [See\,Fig.\ref{DensityDroplet_DFT}]. From the density profiles, a liquid/gas interface is obtained as depicted in Fig.\ref{DensityDroplet_DFT}. A linear fit with all the points located below 8\AA\,is then used to compute the contact angle [See\,Fig\,\ref{DensityDroplet_DFT}]. Uncertainties are evaluated by the standard deviation measured every 100\,ps for 5 independent runs then an additional factor of two is incorporated to account for error in the method for contact angle extraction. In addition, the duration of the simulation is sufficient to reach equilibration as observed in Fig.\,\ref{Error}.a.

\begin{figure}[h!]
\includegraphics[width=8.6cm]{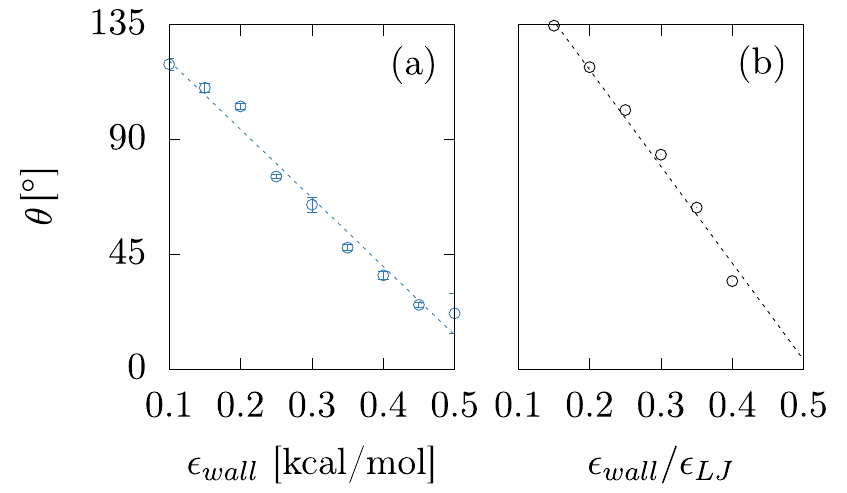}
\caption{Contact angle as a function of the wall energy depth obtained with water/MD (a) and LJ/DFT (b). Dotted lines are obtained through linear fitting.}
\label{ContactAngle}
\end{figure}

\subsubsection{Calculation of the solvent-mediated interactions}
\label{SMI_method}

\begin{figure}[h!]
\includegraphics[width=8.6cm]{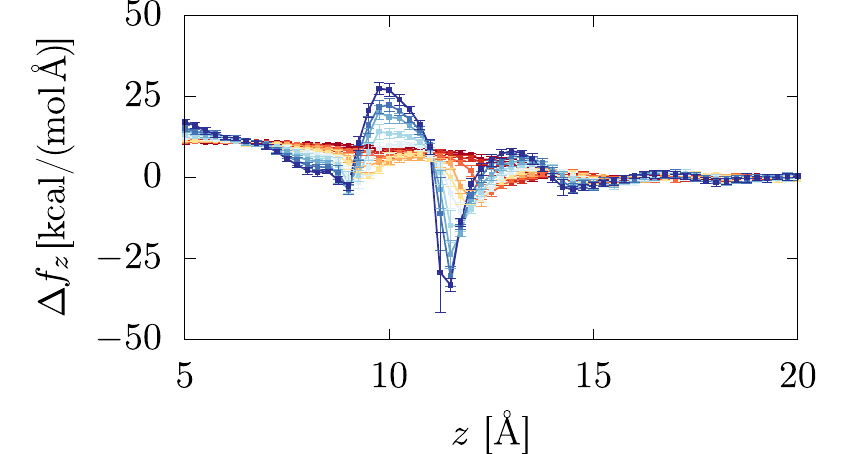}
\caption{Solvophobicity influence on the forces obtained at $a=a_0$ with SPC/E Water. The solvophobicity is increased as coloring goes from blue to red with $\epsilon_{wall}$ going from 0.50\,kcal/mol to 0.05\,kcal/mol. This corresponds to contact angle ranging from $0^\circ$ to $180^\circ$. Each color is separated by $0.05$\,kcal/mol.}
\label{Force}
\end{figure}

\begin{figure}[h!]
\includegraphics[width=8.6cm]{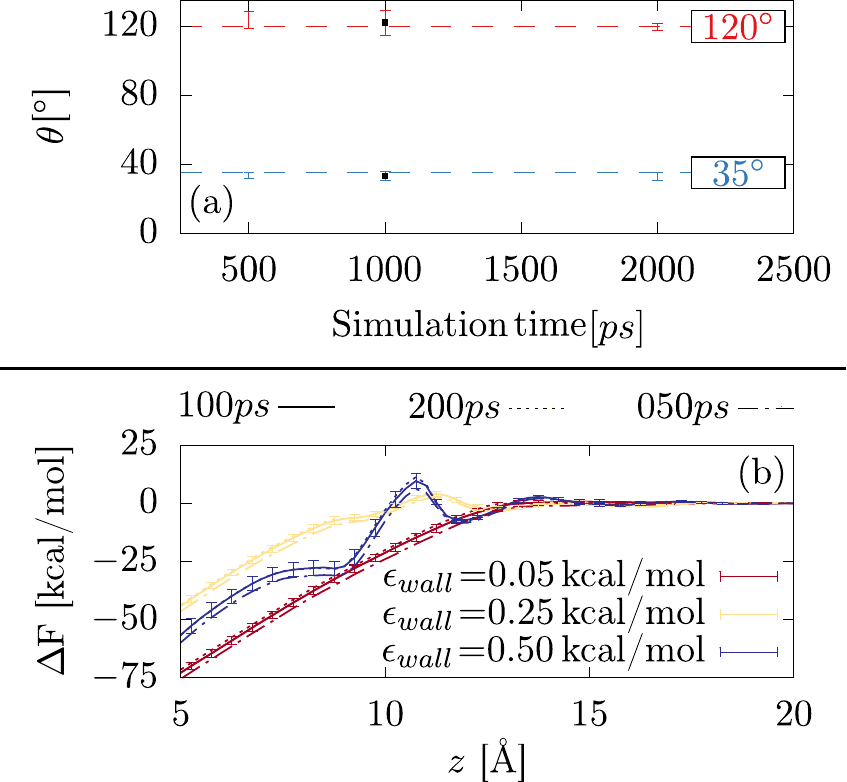}
\caption{Influence of equilibration time over (a) the contact angle and (b) the excess free energy for different degrees of hydrophobicities. For contact angle calculations, red and blue data are obtained with $\epsilon_{wall}$ respectively equal to $0.42$\,kcal/mol and $0.1$\,kcal/mol. The simulation time employed in this work is represented with black dots (1000\,ps for contact angle) and with  plain lines (100\,ps for free energy calculation).}
\label{Error}
\end{figure}

Two nanoslabs which are made of $40+41+40=121\,\textrm{atoms}$ are positioned parallel to each other. The entire simulation box measures $52$\,\AA$\times 52$\,\AA$\times 60$\,\AA\,and contains $5344$ water molecules [See Fig.\,\ref{ProfileSumary_MD}].  The solids are first disposed on top of each other and NVT simulations are performed at $300$\,K during $60$\,ps with a timestep equal to $2$\,fs. After this equilibration procedure, the solids are instantaneously  moved apart by $0.25$\,\AA\,with the distance measured as the difference in height between the center of mass of both slabs. For each separation denoted $z$, the system is equilibrated for $50$\,ps and production run is done during another $50$\,ps. The free energy as a function of $z$ is then given by  numerical integration of the forces:
\label{force}
\begin{equation}
\Delta F(z) =  \int^{z}_\infty \frac{\partial F }{\partial z'}  \mathrm{d}z' = \int^{z}_\infty \left\langle   \vec{f_1}.\vec{u_z}  -\vec{f_2}.\vec{u_z}  \right\rangle  \mathrm{d}z'
\end{equation}
where $\vec{f}_1$ and $\vec{f}_2$  are the forces between water molecules and solid atoms with $1$ and $2$ designating respectively the upper and the lower solids. $\vec{u_z}$ is a unit vector along the $z$ direction going upward. The difference in forces used in the integration scheme Eqn.\,\ref{force} is shown in Fig\,\ref{Force}. Duration of the simulation time is considered sufficient to reach equilibration as assessed by Fig.\,\ref{Error}.b. Error bars in this figure are computed as the standard deviation obtained with 5 independent runs.

\subsection{Density functional theory calculation of Lennard-Jones}

For this second method, liquid particles interact via a LJ potential with $\epsilon_{LJ}$ and $\sigma_{LJ}$ as energy and length parameters. The cutoff distance is equal to\,$3\sigma_{LJ}$. The density and the temperature are respectively $\rho_{LJ}=0.7\,\sigma_{LJ}^{-3}$ and $k_BT=0.8\epsilon_{LJ}$ which is located between the triple point and the critical temperature. This corresponds to the liquid density for a chemical potential supersaturation equal to $\Delta \mu=0.27 k_BT$\cite{Lam2017}. While the value of $\Delta \mu$ influences quantitatively the solvent-mediated forces\cite{Chacko2017}, the supersaturation is chosen in this work to match the ratio of pressure between water coexistence pressure and atmospheric pressure ($P_{coex}=1/20 P_ {atm}$).  Interactions between liquid molecules and solid atoms are also modeled with LJ potential truncated at\,$3\sigma_{LJ}$. Within the DFT framework, free energy is expressed as a functional of the liquid density. For LJ interaction, the potential is separated in two parts, the repulsive part modeled with the White Bear functional \cite{Roth2002Nov} and the attractive part treated in  mean field. The density is computed on a discretized three dimensional grid with 8 lattice points per unit of $\sigma_{LJ}$ and the free energy is obtained through minimization with respect to the density field. In order to match MD calculations that are made in the NVT ensemble, DFT calculations are also run with a fixed number of particles rather than a fixed chemical potential. The DFT method is described in greater details in our previous contributions\cite{Lam2017,Lutsko2010Jan,Lutsko2018May}.  Accuracy of the DFT treatment is discussed in this review\cite{Lutsko2010Jan}. From there, droplet equilibration results were taken from our previous work\cite{Lam2017}. For solvent-mediated interactions, we used the same system as with molecular dynamics simulation of water except that there is no equilibration protocol and the free energy is obtained directly through DFT.

In recent works regarding solvent mediated forces, calculations are performed in $\mu$VT\cite{Chacko2017} or NPT\cite{Jabes2016Aug,Huang2001} ensembles in order to supply particles during the drying transition. To evaluate if our system is large enough to cope with this issue, we also ran calculations in the $\mu$VT with $\Delta \mu=0.27 k_BT$ [See Fig.\,\ref{VsE}.b]. Results are not significantly different from those obtained in NVT thus justifying our approach.

\section{Results and discussion}

\begin{figure}[h!]
\includegraphics[width=8.6cm]{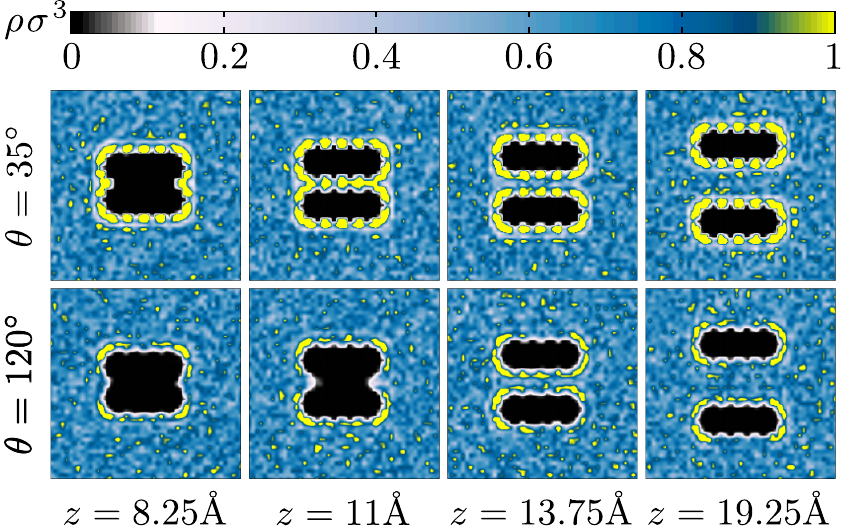}
\caption{Liquid density profiles for water molecules confined between two nanoslabs obtained for $a=a_0$ at different distances and degrees of solvophobicity. Contact angles of $35^\circ$ and $120^\circ$ are obtained  with $\epsilon_{wall}$ respectively equal to $0.42$\,kcal/mol and $0.10$\,kcal/mol. Each image is a slice of width $1$\AA\,passing through the middle of the solute and measuring $48$\AA$\times48$\AA.}
\label{ProfileSumary_MD}
\end{figure}

\begin{figure}[h!]
\includegraphics[width=8.6cm]{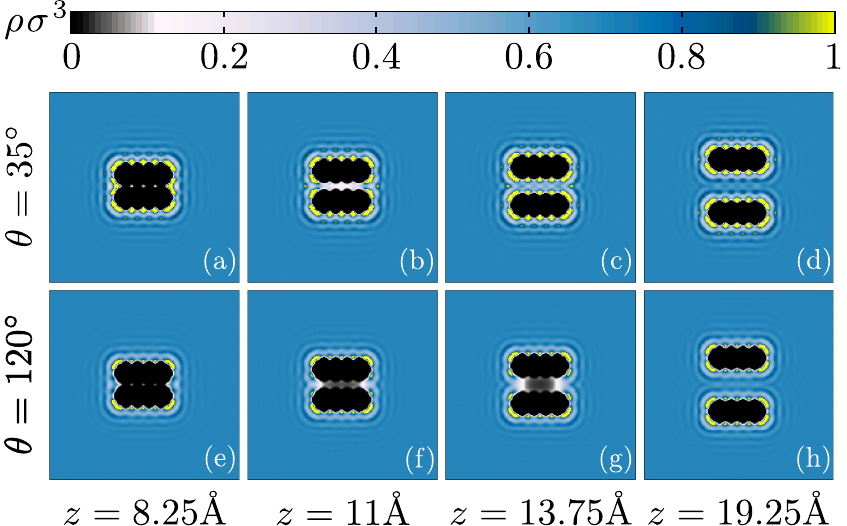}
\caption{Liquid density profiles for Lennard-Jones particles confined between two nanoslabs obtained for $a=a_0$ at different distances and degrees of solvophobicity. Contact angles of $35^\circ$ and $120^\circ$ are obtained  with $\epsilon_{wall}$ respectively equal to $0.4\,\epsilon_{LJ}$ and $0.2\,\epsilon_{LJ}$. Each image is a slice of width $0.275$\AA\,passing through the middle of the solute and measuring $68.75$\AA$\times 68.75$\AA.}
\label{ProfileSumary_DFT}
\end{figure}

Figures \,\ref{ProfileSumary_MD} and \ref{ProfileSumary_DFT} show typical results obtained respectively for water and Lennard-Jones. In both cases, when walls are solvophilic, the gap between the two slabs is filled with liquid even at small distances [See\,Figs.\,(\ref{ProfileSumary_MD}b,\,\ref{ProfileSumary_DFT}b)]. For solvophobic walls, this happens only for large enough separations. In addition, structuring can be observed in the vicinity of the slabs especially when looking at the density profiles obtained by DFT of Lennard-Jones particles [See Fig. \,\ref{ProfileSumary_DFT}]. The structure is more pronounced for solvophilic walls.

\subsection{Influence of the wall lattice spacing}

\begin{figure}[h!]
\includegraphics[width=8.6cm]{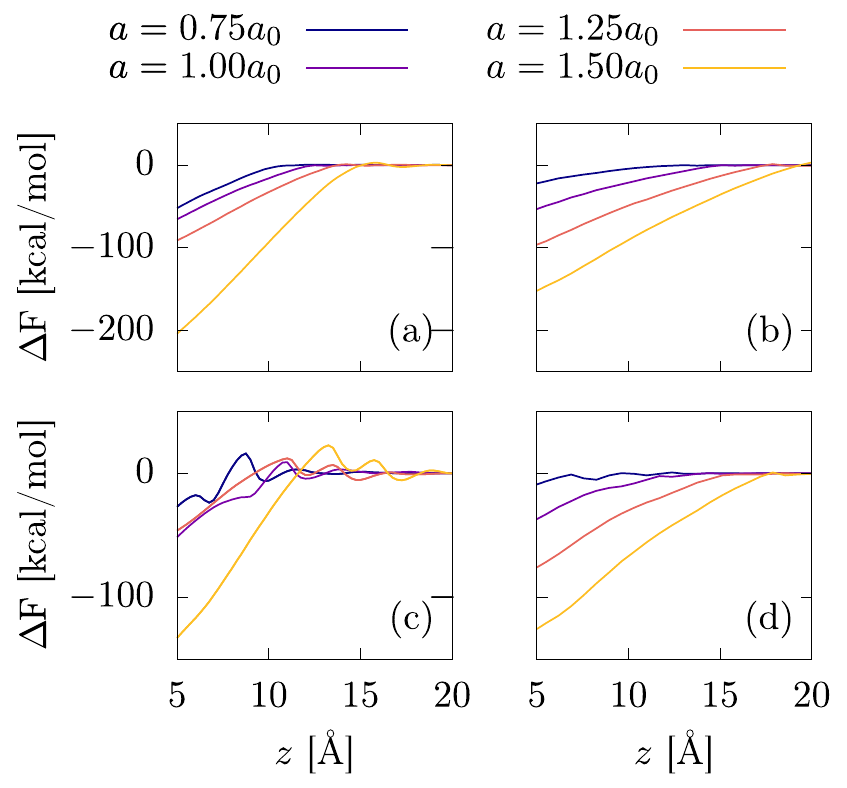}
\caption{Lattice spacing influence on the excess free energy obtained with SPC/E Water (left) and with Lennard-Jones (right). Calculations are run at a value of $\epsilon_{wall}$ for which the contact angle is  $\theta=$  120$^\circ$  (top) and at $\theta=$ 35$^\circ$ (bottom) for $a=a_0$. Values of $\epsilon_{wall}$ are given in captions of Fig.\ref{ProfileSumary_MD} and Fig.\ref{ProfileSumary_DFT}.}
\label{VsLattice}
\end{figure}

In this first study, we worked at a fixed value of $\epsilon_{wall}$ while changing the wall lattice spacing so that both the structure and the hydrophobicity are modified. In Figs\,.\,(\ref{VsLattice}a,\,\ref{VsLattice}b), excess free energy is plotted for a moderate value of $\epsilon_{wall}$ ($0.1$\,kcal/mol and $0.2\epsilon_{LJ}$) that in both cases lead to 35$^\circ$ when $a=a_0$. An almost linear decrease is observed which is consistent with previous works on solvophobic attraction\cite{Lam2017,Jabes2016Aug,Li2005}. In particular, near contact, one can show that the slope depends solely on bulk properties using a capillary model\cite{Chacko2017}. At intermediate distances, the slope is also influenced by the wall solvophobicity since the presence of a meniscus leads to a non trivial shape of the gaseous phase\cite{Chacko2017,Lam2017}. When $a$ is reduced, the walls are denser and thus becomes less solvophobic. Therefore, both the range and the height of the solvent-mediated interaction are reduced. The results obtained with LJ DFT and with water MD are qualitatively similar and we demonstrate that in this case, LJ can be used to reproduce water-mediated interactions. For solvophilic walls, the results of the comparison are not so close [See Figs\,.\,(\ref{VsLattice}c,\,\ref{VsLattice}d)]. In both systems, oscillations in the free energy are observed due to layering of the liquid near the walls. However, the oscillation amplitudes vary significantly between water and the LJ fluid. This is likely due to the asymmetry of water molecules: as they pack together to form denser layers near the wall, their interlayer distance does not depend solely on their average size but rather on their size in some particular directions\cite{Banerjee2015}. 

\subsection{Influence of the wall energy}

\begin{figure}[h!]
\includegraphics[width=8.6cm]{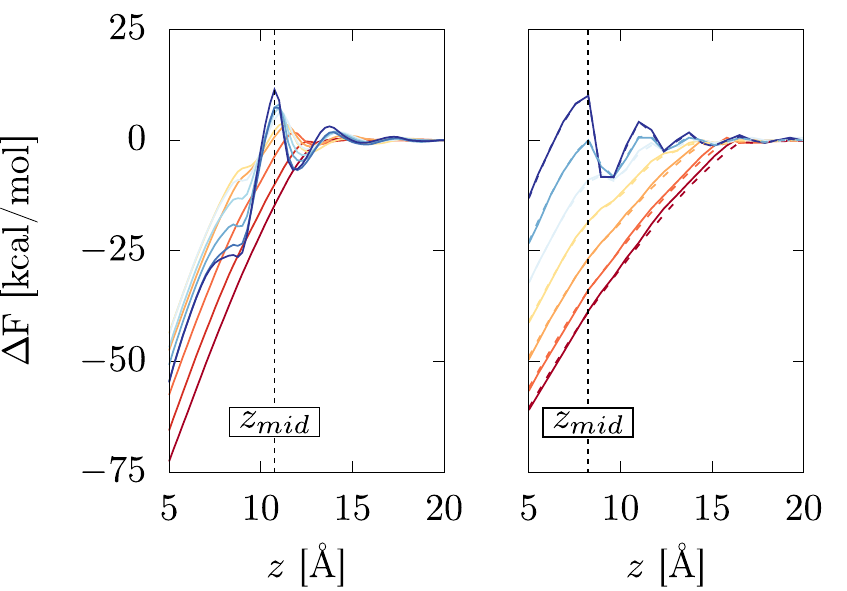}
\caption{Wall energy influence on the excess free energy obtained at $a=a_0$ with SPC/E Water (a) and with Lennard-Jones (b). The solvophobicity is increased as coloring goes from blue to red with $\epsilon_{wall}$ going from 0.50\,kcal/mol ($0.65\epsilon_{LJ}$) to 0.05\,kcal/mol  ($0.05\epsilon_{LJ}$) for water (for LJ). This corresponds to contact angle ranging from $0^\circ$ to $180^\circ$. Each color is separated by $0.05$\,kcal/mol (by\,$0.1\epsilon_{LJ}$) for water (for LJ). Dash lines correspond to results obtained with LJ/DFT in the $\mu$VT ensemble using $\Delta \mu =0.27 k_BT$.}
\label{VsE}
\end{figure}

Solvent-mediated interactions are plotted at a fixed value of $a=a_0$ but for different values of $\epsilon_{wall}$ in Fig.\,\ref{VsE}. When comparing results from MD water and DFT LJ, several similarities can be identified. First, when the walls are solvophobic,  free energy monotonically increases as the nanoslabs are pulled apart with an almost linear behavior. Then, when the walls are solvophilic, damped oscillations are observed because of the emergence of structured layers near the wall. Also, the lowest energy state is always at contact meaning that the nanoparticle would preferentially stay near the wall as long as it overcomes the intermediate free energy barrier. Finally, the range of the depletion force does not go beyond $20$\,\AA\,which corresponds to approximately 7 liquid layers. These similarities were already raised in the literature\cite{Ashbaugh2013Aug,Evans2015Apr,Evans2015Jul} and our work allows for a more direct comparison as we studied the same system (ie. two nanoslabs made of the same structure) while only changing the liquid nature. 

\begin{figure}[h!]
\includegraphics[width=8.6cm]{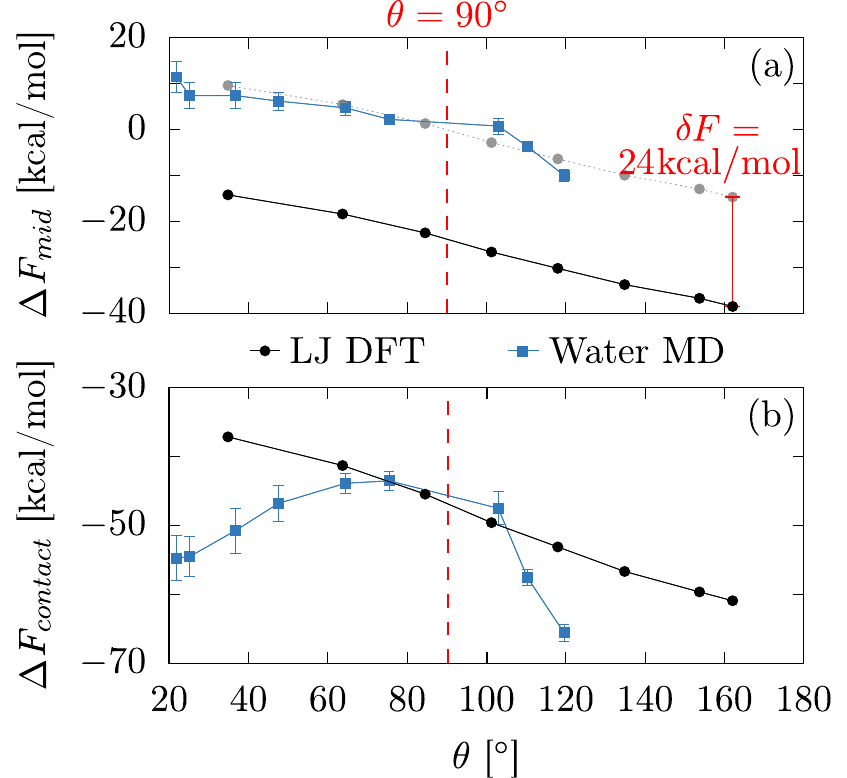}
\caption{Comparison between results of LJ DFT and water MD using the contact angle dependence of $\Delta F_{mid}$ (a) and $\Delta F_{contact}$ (b). In grey, results for $\Delta F_{mid}$ obtained with LJ DFT is vertically shifted in order to show that the difference with water MD is only a constant.}
\label{MDvsDFT}
\end{figure}

In order to quantitatively compare results from water MD and LJ DFT, we define two positions: (i)\,$z=5$\AA\,gives $\Delta F_{contact}$ and (ii) the position, denoted $z_{mid}$, at which the most solvophilic interaction reaches its maximum is used as an intermediate value called $\Delta F_{mid}$ [See Fig.\,\ref{VsE}]. In Fig.\ref{MDvsDFT}, results are reported for different contact angles that are determined after the equilibration of sessile drops. For the highest degrees of solvophilicity (and solvophobicity), droplets are not stable and the contact angle is trivially $0^\circ$ (and $180^\circ$). Therefore, when reporting $\Delta F_{mid}$  and $\Delta F_{contact}$  as a function of the corresponding contact angle, not all the data from Fig.\,\ref{VsE} are considered. As the contact angle is increased, $\Delta F_{mid}$ decreases almost linearly. Water MD and LJ DFT curves have similar slopes and the constant difference between the two curves is roughly of $24$\,kcal/mol. However, the sign of $\Delta F_{mid}$ is different which indicates that qualitatively different behaviors are expected. In the water case, this intermediate distance is less energetically favorable than having the nanoslabs far from each other. For $\Delta F_{contact}$, LJ DFT and water MD also lead to qualitatively different results. Indeed, while for LJ, $\Delta F_{contact}$, like $\Delta F_{mid}$, decreases linearly, for water, $\Delta F_{contact}$ is non-monotonic and peaks around $80^\circ$. As already raised in the previous section, LJ is not well-adapted to model water at contact because water has orientational order especially for solvophilic walls which can not be seen with LJ. Furthermore, while there is an intermediate range of solvophibicity ($\theta \in [70^\circ:110^\circ]$) where a good agreement for $\Delta F_{contact}$ is found, the signs of $\Delta F_{mid}$ are different as raised above.

\subsection{Hysteresis in solvent-mediated forces}

\begin{figure}[h!]
\includegraphics[width=8.6cm]{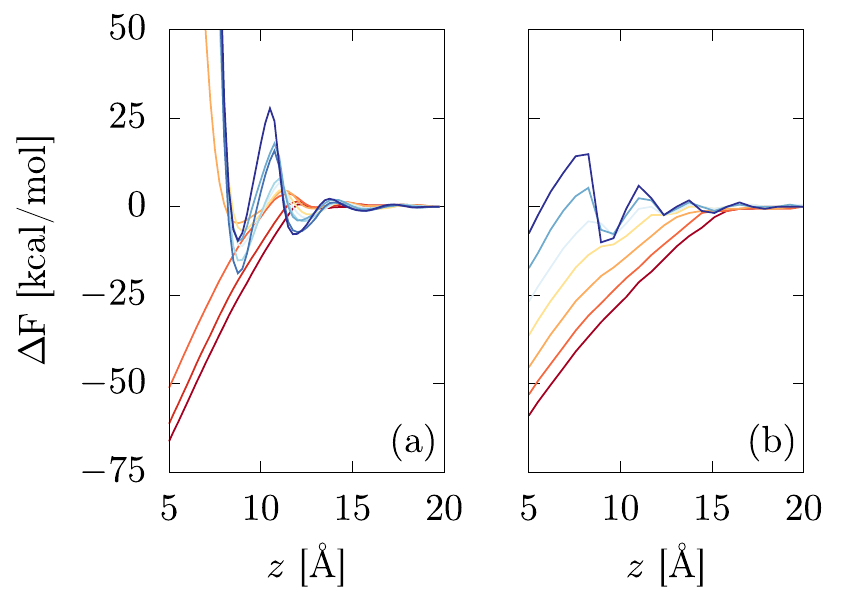}
\caption{Solvent-mediated forces at $a=a_0$ with SPC/E Water (a) and with Lennard-Jones (b) obtained as the solutes are brought closer to each other. Color designations are described in Fig.\,\ref{VsE}.}
\label{UpVsDown}
\end{figure}

Throughout this study, results on the solvent-mediated forces were obtained as the two solutes are pulled apart from each other. Yet, another possibility concerns the case when the slabs are disposed far from each other and then brought together. This leads to the question of reversibility of the interaction. In Fig.\,\ref{UpVsDown}, the solvent mediated forces are plotted in this second approach. In the case of water MD, qualitative agreement is found when comparing results from Fig.\,\ref{VsE} with solvophobic walls. However, strong repulsive interactions are observed with solvophilic walls.  This results from water molecules that can be trapped between the two plates if they are brought together too rapidly. Furthermore, the capillary evaporation which is not observed when the solutes are pulled apart, is not only driven by the solute interdistance and additional order parameters such as the solvent density between the solutes can be used \cite{Altabet2017Mar,Remsing2015Jul,Bolhuis2000Nov}. Essentially, the time for gas to nucleate between the two walls is so large that we can not obtain the equilibrium state with brute-force molecular dynamics simulations\cite{Bolhuis2000Nov}. This apparent hysteresis is not found in LJ DFT since the technique enables to circumvent any of these kinetic issues and leads directly to the most stable state in which the gap between the walls is emptied of liquid.

\section{Conclusions}

In summary, solvent-mediated forces were measured in a very generic case of two nanoslabs embedded in a liquid. Two different models for the liquid along with two different methods for measuring free energy were employed. Such a direct comparison of these two approaches allowed us to identify both similarities and differences. On the one hand, oscillations for solvophilic walls and a linear decrease for solvophobic walls were observed with the two liquids. In addition, the range of the depletion force and the presence of a minimum at contact are two additional features that seem to support the idea of a universal behavior of the solvent-mediated forces. On the other hand, no region of solvophobicity seems to show a quantitative agreement between water and LJ. In particular, amplitudes of the oscillations and the resulting sign of the free energy for intermediate distances are different. Also, the value of the free energy at contact does not have the same behavior as the contact angle is changed. Ultimately, using LJ or water in order to model solvent mediated forces should depend on the desired level of accuracy and our results provide a benchmark that quantifies the error made if one wishes to use LJ instead of water.

\section{Acknowledgement}
The work of JL was funded by the European Union's Horizon 2020 research and innovation program  within the AMECRYS project under grant agreement no. 712965. JFL thanks the European Space Agency (ESA) and the Belgian Federal Science
Policy Office (BELSPO) for their support in the framework of the PRODEX Programme, contract number ESA17
AO-2004-070. Computational resources have been provided by the Consortium des Equipements de Calcul Intensif (CECI) and by the Fédération Lyonnaise de Modélisation et Sciences Numériques (FLMSN).

\bibliography{biblio}

\begin{thebibliography}{65}%
\makeatletter
\providecommand \@ifxundefined [1]{%
 \@ifx{#1\undefined}
}%
\providecommand \@ifnum [1]{%
 \ifnum #1\expandafter \@firstoftwo
 \else \expandafter \@secondoftwo
 \fi
}%
\providecommand \@ifx [1]{%
 \ifx #1\expandafter \@firstoftwo
 \else \expandafter \@secondoftwo
 \fi
}%
\providecommand \natexlab [1]{#1}%
\providecommand \enquote  [1]{``#1''}%
\providecommand \bibnamefont  [1]{#1}%
\providecommand \bibfnamefont [1]{#1}%
\providecommand \citenamefont [1]{#1}%
\providecommand \href@noop [0]{\@secondoftwo}%
\providecommand \href [0]{\begingroup \@sanitize@url \@href}%
\providecommand \@href[1]{\@@startlink{#1}\@@href}%
\providecommand \@@href[1]{\endgroup#1\@@endlink}%
\providecommand \@sanitize@url [0]{\catcode `\\12\catcode `\$12\catcode
  `\&12\catcode `\#12\catcode `\^12\catcode `\_12\catcode `\%12\relax}%
\providecommand \@@startlink[1]{}%
\providecommand \@@endlink[0]{}%
\providecommand \url  [0]{\begingroup\@sanitize@url \@url }%
\providecommand \@url [1]{\endgroup\@href {#1}{\urlprefix }}%
\providecommand \urlprefix  [0]{URL }%
\providecommand \Eprint [0]{\href }%
\providecommand \doibase [0]{http://dx.doi.org/}%
\providecommand \selectlanguage [0]{\@gobble}%
\providecommand \bibinfo  [0]{\@secondoftwo}%
\providecommand \bibfield  [0]{\@secondoftwo}%
\providecommand \translation [1]{[#1]}%
\providecommand \BibitemOpen [0]{}%
\providecommand \bibitemStop [0]{}%
\providecommand \bibitemNoStop [0]{.\EOS\space}%
\providecommand \EOS [0]{\spacefactor3000\relax}%
\providecommand \BibitemShut  [1]{\csname bibitem#1\endcsname}%
\let\auto@bib@innerbib\@empty
\bibitem [{\citenamefont {Leikin}, \citenamefont {Rau},\ and\ \citenamefont
  {Parsegian}(1994)}]{Leikin1994}%
  \BibitemOpen
  \bibfield  {author} {\bibinfo {author} {\bibfnamefont {S.}~\bibnamefont
  {Leikin}}, \bibinfo {author} {\bibfnamefont {D.~C.}\ \bibnamefont {Rau}}, \
  and\ \bibinfo {author} {\bibfnamefont {V.~A.}\ \bibnamefont {Parsegian}},\
  }\href@noop {} {\bibfield  {journal} {\bibinfo  {journal} {Proc. Natl. Acad.
  Sci. U.S.A.}\ }\textbf {\bibinfo {volume} {91}},\ \bibinfo {pages} {276}
  (\bibinfo {year} {1994})}\BibitemShut {NoStop}%
\bibitem [{\citenamefont {Chandler}(2005)}]{Chandler2005}%
  \BibitemOpen
  \bibfield  {author} {\bibinfo {author} {\bibfnamefont {D.}~\bibnamefont
  {Chandler}},\ }\href {\doibase 10.1038/nature04162} {\bibfield  {journal}
  {\bibinfo  {journal} {Nature}\ }\textbf {\bibinfo {volume} {437}},\ \bibinfo
  {pages} {640} (\bibinfo {year} {2005})}\BibitemShut {NoStop}%
\bibitem [{\citenamefont {Kowalik}\ \emph {et~al.}(2017)\citenamefont
  {Kowalik}, \citenamefont {Schlaich}, \citenamefont {Kandu{\ifmmode \check{c}
  \else \v{c}\fi}}, \citenamefont {Schneck},\ and\ \citenamefont
  {Netz}}]{Kowalik2017}%
  \BibitemOpen
  \bibfield  {author} {\bibinfo {author} {\bibfnamefont {B.}~\bibnamefont
  {Kowalik}}, \bibinfo {author} {\bibfnamefont {A.}~\bibnamefont {Schlaich}},
  \bibinfo {author} {\bibfnamefont {M.}~\bibnamefont {Kandu{\ifmmode \check{c}
  \else \v{c}\fi}}}, \bibinfo {author} {\bibfnamefont {E.}~\bibnamefont
  {Schneck}}, \ and\ \bibinfo {author} {\bibfnamefont {R.~R.}\ \bibnamefont
  {Netz}},\ }\href {\doibase 10.1021/acs.jpclett.7b00977} {\bibfield  {journal}
  {\bibinfo  {journal} {J. Phys. Chem. Lett.}\ }\textbf {\bibinfo {volume}
  {8}},\ \bibinfo {pages} {2869} (\bibinfo {year} {2017})}\BibitemShut
  {NoStop}%
\bibitem [{\citenamefont {Maibaum}, \citenamefont {Dinner},\ and\ \citenamefont
  {Chandler}(2004)}]{Maibaum2004}%
  \BibitemOpen
  \bibfield  {author} {\bibinfo {author} {\bibfnamefont {L.}~\bibnamefont
  {Maibaum}}, \bibinfo {author} {\bibfnamefont {A.~R.}\ \bibnamefont {Dinner}},
  \ and\ \bibinfo {author} {\bibfnamefont {D.}~\bibnamefont {Chandler}},\
  }\href {\doibase 10.1021/jp037487t} {\bibfield  {journal} {\bibinfo
  {journal} {J. Phys. Chem. B}\ }\textbf {\bibinfo {volume} {108}},\ \bibinfo
  {pages} {6778} (\bibinfo {year} {2004})}\BibitemShut {NoStop}%
\bibitem [{\citenamefont {Tiwary}\ \emph
  {et~al.}(2015{\natexlab{a}})\citenamefont {Tiwary}, \citenamefont {Mondal},
  \citenamefont {Morrone},\ and\ \citenamefont {Berne}}]{Tiwary2015a}%
  \BibitemOpen
  \bibfield  {author} {\bibinfo {author} {\bibfnamefont {P.}~\bibnamefont
  {Tiwary}}, \bibinfo {author} {\bibfnamefont {J.}~\bibnamefont {Mondal}},
  \bibinfo {author} {\bibfnamefont {J.~A.}\ \bibnamefont {Morrone}}, \ and\
  \bibinfo {author} {\bibfnamefont {B.~J.}\ \bibnamefont {Berne}},\ }\href
  {\doibase 10.1073/pnas.1516652112} {\bibfield  {journal} {\bibinfo  {journal}
  {Proc. Natl. Acad. Sci. U.S.A.}\ }\textbf {\bibinfo {volume} {112}},\
  \bibinfo {pages} {12015} (\bibinfo {year} {2015}{\natexlab{a}})}\BibitemShut
  {NoStop}%
\bibitem [{\citenamefont {Tiwary}\ \emph
  {et~al.}(2015{\natexlab{b}})\citenamefont {Tiwary}, \citenamefont
  {Limongelli}, \citenamefont {Salvalaglio},\ and\ \citenamefont
  {Parrinello}}]{Tiwary2015b}%
  \BibitemOpen
  \bibfield  {author} {\bibinfo {author} {\bibfnamefont {P.}~\bibnamefont
  {Tiwary}}, \bibinfo {author} {\bibfnamefont {V.}~\bibnamefont {Limongelli}},
  \bibinfo {author} {\bibfnamefont {M.}~\bibnamefont {Salvalaglio}}, \ and\
  \bibinfo {author} {\bibfnamefont {M.}~\bibnamefont {Parrinello}},\ }\href
  {\doibase 10.1073/pnas.1424461112} {\bibfield  {journal} {\bibinfo  {journal}
  {Proc. Natl. Acad. Sci. U.S.A.}\ }\textbf {\bibinfo {volume} {112}},\
  \bibinfo {pages} {E386} (\bibinfo {year} {2015}{\natexlab{b}})}\BibitemShut
  {NoStop}%
\bibitem [{\citenamefont {Jamadagni}, \citenamefont {Godawat},\ and\
  \citenamefont {Garde}(2009)}]{Jamadagni2009}%
  \BibitemOpen
  \bibfield  {author} {\bibinfo {author} {\bibfnamefont {S.~N.}\ \bibnamefont
  {Jamadagni}}, \bibinfo {author} {\bibfnamefont {R.}~\bibnamefont {Godawat}},
  \ and\ \bibinfo {author} {\bibfnamefont {S.}~\bibnamefont {Garde}},\ }\href
  {\doibase 10.1021/la9011839} {\bibfield  {journal} {\bibinfo  {journal}
  {Langmuir}\ }\textbf {\bibinfo {volume} {25}},\ \bibinfo {pages} {13092}
  (\bibinfo {year} {2009})}\BibitemShut {NoStop}%
\bibitem [{\citenamefont {Camilloni}\ \emph {et~al.}(2016)\citenamefont
  {Camilloni}, \citenamefont {Bonetti}, \citenamefont {Morrone}, \citenamefont
  {Giri}, \citenamefont {Dobson}, \citenamefont {Brunori}, \citenamefont
  {Gianni},\ and\ \citenamefont {Vendruscolo}}]{Camilloni2016}%
  \BibitemOpen
  \bibfield  {author} {\bibinfo {author} {\bibfnamefont {C.}~\bibnamefont
  {Camilloni}}, \bibinfo {author} {\bibfnamefont {D.}~\bibnamefont {Bonetti}},
  \bibinfo {author} {\bibfnamefont {A.}~\bibnamefont {Morrone}}, \bibinfo
  {author} {\bibfnamefont {R.}~\bibnamefont {Giri}}, \bibinfo {author}
  {\bibfnamefont {C.~M.}\ \bibnamefont {Dobson}}, \bibinfo {author}
  {\bibfnamefont {M.}~\bibnamefont {Brunori}}, \bibinfo {author} {\bibfnamefont
  {S.}~\bibnamefont {Gianni}}, \ and\ \bibinfo {author} {\bibfnamefont
  {M.}~\bibnamefont {Vendruscolo}},\ }\href {\doibase 10.1038/srep28285}
  {\bibfield  {journal} {\bibinfo  {journal} {Sci. Rep.}\ }\textbf {\bibinfo
  {volume} {6}},\ \bibinfo {pages} {28285} (\bibinfo {year}
  {2016})}\BibitemShut {NoStop}%
\bibitem [{\citenamefont {Spolar}, \citenamefont {Ha},\ and\ \citenamefont
  {Record}(1989)}]{Spolar1989}%
  \BibitemOpen
  \bibfield  {author} {\bibinfo {author} {\bibfnamefont {R.~S.}\ \bibnamefont
  {Spolar}}, \bibinfo {author} {\bibfnamefont {J.~H.}\ \bibnamefont {Ha}}, \
  and\ \bibinfo {author} {\bibfnamefont {M.~T.}\ \bibnamefont {Record}},\
  }\href@noop {} {\bibfield  {journal} {\bibinfo  {journal} {Proc. Natl. Acad.
  Sci. U.S.A.}\ }\textbf {\bibinfo {volume} {86}},\ \bibinfo {pages} {8382}
  (\bibinfo {year} {1989})}\BibitemShut {NoStop}%
\bibitem [{\citenamefont {Li}, \citenamefont {Bedrov},\ and\ \citenamefont
  {Smith}(2005)}]{Li2005}%
  \BibitemOpen
  \bibfield  {author} {\bibinfo {author} {\bibfnamefont {L.}~\bibnamefont
  {Li}}, \bibinfo {author} {\bibfnamefont {D.}~\bibnamefont {Bedrov}}, \ and\
  \bibinfo {author} {\bibfnamefont {G.~D.}\ \bibnamefont {Smith}},\ }\href
  {\doibase 10.1103/PhysRevE.71.011502} {\bibfield  {journal} {\bibinfo
  {journal} {Phys. Rev. E}\ }\textbf {\bibinfo {volume} {71}},\ \bibinfo
  {pages} {011502} (\bibinfo {year} {2005})}\BibitemShut {NoStop}%
\bibitem [{\citenamefont {Patel}, \citenamefont {Varilly},\ and\ \citenamefont
  {Chandler}(2010)}]{Patel2010}%
  \BibitemOpen
  \bibfield  {author} {\bibinfo {author} {\bibfnamefont {A.~J.}\ \bibnamefont
  {Patel}}, \bibinfo {author} {\bibfnamefont {P.}~\bibnamefont {Varilly}}, \
  and\ \bibinfo {author} {\bibfnamefont {D.}~\bibnamefont {Chandler}},\ }\href
  {\doibase 10.1021/jp909048f} {\bibfield  {journal} {\bibinfo  {journal} {J.
  Phys. Chem. B}\ }\textbf {\bibinfo {volume} {114}},\ \bibinfo {pages} {1632}
  (\bibinfo {year} {2010})}\BibitemShut {NoStop}%
\bibitem [{\citenamefont {Patel}\ \emph {et~al.}(2011)\citenamefont {Patel},
  \citenamefont {Varilly}, \citenamefont {Jamadagni}, \citenamefont {Acharya},
  \citenamefont {Garde},\ and\ \citenamefont {Chandler}}]{Patel2011}%
  \BibitemOpen
  \bibfield  {author} {\bibinfo {author} {\bibfnamefont {A.~J.}\ \bibnamefont
  {Patel}}, \bibinfo {author} {\bibfnamefont {P.}~\bibnamefont {Varilly}},
  \bibinfo {author} {\bibfnamefont {S.~N.}\ \bibnamefont {Jamadagni}}, \bibinfo
  {author} {\bibfnamefont {H.}~\bibnamefont {Acharya}}, \bibinfo {author}
  {\bibfnamefont {S.}~\bibnamefont {Garde}}, \ and\ \bibinfo {author}
  {\bibfnamefont {D.}~\bibnamefont {Chandler}},\ }\href@noop {} {\bibfield
  {journal} {\bibinfo  {journal} {Proceedings of the National Academy of
  Sciences}\ }\textbf {\bibinfo {volume} {108}},\ \bibinfo {pages} {17678}
  (\bibinfo {year} {2011})}\BibitemShut {NoStop}%
\bibitem [{\citenamefont {Huang}, \citenamefont {Geissler},\ and\ \citenamefont
  {Chandler}(2001)}]{Huang2001}%
  \BibitemOpen
  \bibfield  {author} {\bibinfo {author} {\bibfnamefont {D.~M.}\ \bibnamefont
  {Huang}}, \bibinfo {author} {\bibfnamefont {P.~L.}\ \bibnamefont {Geissler}},
  \ and\ \bibinfo {author} {\bibfnamefont {D.}~\bibnamefont {Chandler}},\
  }\href {\doibase 10.1021/jp0104029} {\bibfield  {journal} {\bibinfo
  {journal} {J. Phys. Chem. B}\ }\textbf {\bibinfo {volume} {105}},\ \bibinfo
  {pages} {6704} (\bibinfo {year} {2001})}\BibitemShut {NoStop}%
\bibitem [{\citenamefont {Banerjee}, \citenamefont {Singh},\ and\ \citenamefont
  {Bagchi}(2015)}]{Banerjee2015}%
  \BibitemOpen
  \bibfield  {author} {\bibinfo {author} {\bibfnamefont {S.}~\bibnamefont
  {Banerjee}}, \bibinfo {author} {\bibfnamefont {R.~S.}\ \bibnamefont {Singh}},
  \ and\ \bibinfo {author} {\bibfnamefont {B.}~\bibnamefont {Bagchi}},\ }\href
  {\doibase 10.1063/1.4916744} {\bibfield  {journal} {\bibinfo  {journal} {J.
  Chem. Phys.}\ }\textbf {\bibinfo {volume} {142}},\ \bibinfo {pages} {134505}
  (\bibinfo {year} {2015})}\BibitemShut {NoStop}%
\bibitem [{\citenamefont {Kandu{\ifmmode \check{c} \else \v{c}\fi}}\ and\
  \citenamefont {Netz}(2015)}]{Kanduc2015}%
  \BibitemOpen
  \bibfield  {author} {\bibinfo {author} {\bibfnamefont {M.}~\bibnamefont
  {Kandu{\ifmmode \check{c} \else \v{c}\fi}}}\ and\ \bibinfo {author}
  {\bibfnamefont {R.~R.}\ \bibnamefont {Netz}},\ }\href {\doibase
  10.1073/pnas.1504919112} {\bibfield  {journal} {\bibinfo  {journal} {Proc.
  Natl. Acad. Sci. U.S.A.}\ }\textbf {\bibinfo {volume} {112}},\ \bibinfo
  {pages} {12338} (\bibinfo {year} {2015})}\BibitemShut {NoStop}%
\bibitem [{\citenamefont {Kandu{\ifmmode \check{c} \else \v{c}\fi}}\ \emph
  {et~al.}(2016)\citenamefont {Kandu{\ifmmode \check{c} \else \v{c}\fi}},
  \citenamefont {Schlaich}, \citenamefont {Schneck},\ and\ \citenamefont
  {Netz}}]{Kanduc2016}%
  \BibitemOpen
  \bibfield  {author} {\bibinfo {author} {\bibfnamefont {M.}~\bibnamefont
  {Kandu{\ifmmode \check{c} \else \v{c}\fi}}}, \bibinfo {author} {\bibfnamefont
  {A.}~\bibnamefont {Schlaich}}, \bibinfo {author} {\bibfnamefont
  {E.}~\bibnamefont {Schneck}}, \ and\ \bibinfo {author} {\bibfnamefont
  {R.~R.}\ \bibnamefont {Netz}},\ }\href {\doibase
  10.1021/acs.langmuir.6b01727} {\bibfield  {journal} {\bibinfo  {journal}
  {Langmuir}\ }\textbf {\bibinfo {volume} {32}},\ \bibinfo {pages} {8767}
  (\bibinfo {year} {2016})}\BibitemShut {NoStop}%
\bibitem [{\citenamefont {Chacko}, \citenamefont {Evans},\ and\ \citenamefont
  {Archer}(2017)}]{Chacko2017}%
  \BibitemOpen
  \bibfield  {author} {\bibinfo {author} {\bibfnamefont {B.}~\bibnamefont
  {Chacko}}, \bibinfo {author} {\bibfnamefont {R.}~\bibnamefont {Evans}}, \
  and\ \bibinfo {author} {\bibfnamefont {A.~J.}\ \bibnamefont {Archer}},\
  }\href {\doibase 10.1063/1.4978352} {\bibfield  {journal} {\bibinfo
  {journal} {J. Chem. Phys.}\ }\textbf {\bibinfo {volume} {146}},\ \bibinfo
  {pages} {124703} (\bibinfo {year} {2017})}\BibitemShut {NoStop}%
\bibitem [{\citenamefont {Lam}\ and\ \citenamefont {Lutsko}(2017)}]{Lam2017}%
  \BibitemOpen
  \bibfield  {author} {\bibinfo {author} {\bibfnamefont {J.}~\bibnamefont
  {Lam}}\ and\ \bibinfo {author} {\bibfnamefont {J.~F.}\ \bibnamefont
  {Lutsko}},\ }\href {\doibase 10.1039/C7NR07218J} {\bibfield  {journal}
  {\bibinfo  {journal} {Nanoscale}\ }\textbf {\bibinfo {volume} {9}},\ \bibinfo
  {pages} {17099} (\bibinfo {year} {2017})}\BibitemShut {NoStop}%
\bibitem [{\citenamefont {Stock}\ \emph {et~al.}(2017)\citenamefont {Stock},
  \citenamefont {Monroe}, \citenamefont {Utzig}, \citenamefont {Smith},
  \citenamefont {Shell},\ and\ \citenamefont {Valtiner}}]{Stock2017}%
  \BibitemOpen
  \bibfield  {author} {\bibinfo {author} {\bibfnamefont {P.}~\bibnamefont
  {Stock}}, \bibinfo {author} {\bibfnamefont {J.~I.}\ \bibnamefont {Monroe}},
  \bibinfo {author} {\bibfnamefont {T.}~\bibnamefont {Utzig}}, \bibinfo
  {author} {\bibfnamefont {D.~J.}\ \bibnamefont {Smith}}, \bibinfo {author}
  {\bibfnamefont {M.~S.}\ \bibnamefont {Shell}}, \ and\ \bibinfo {author}
  {\bibfnamefont {M.}~\bibnamefont {Valtiner}},\ }\href {\doibase
  10.1021/acsnano.6b06360} {\bibfield  {journal} {\bibinfo  {journal} {ACS
  Nano}\ }\textbf {\bibinfo {volume} {11}},\ \bibinfo {pages} {2586} (\bibinfo
  {year} {2017})}\BibitemShut {NoStop}%
\bibitem [{\citenamefont {Pashley}\ \emph {et~al.}(1985)\citenamefont
  {Pashley}, \citenamefont {McGuiggan}, \citenamefont {Ninham},\ and\
  \citenamefont {Evans}}]{Pashley1985Sep}%
  \BibitemOpen
  \bibfield  {author} {\bibinfo {author} {\bibfnamefont {R.}~\bibnamefont
  {Pashley}}, \bibinfo {author} {\bibfnamefont {P.}~\bibnamefont {McGuiggan}},
  \bibinfo {author} {\bibfnamefont {B.}~\bibnamefont {Ninham}}, \ and\ \bibinfo
  {author} {\bibfnamefont {D.}~\bibnamefont {Evans}},\ }\href {\doibase
  10.1126/science.4035349} {\bibfield  {journal} {\bibinfo  {journal}
  {Science}\ }\textbf {\bibinfo {volume} {229}},\ \bibinfo {pages} {1088}
  (\bibinfo {year} {1985})}\BibitemShut {NoStop}%
\bibitem [{\citenamefont {Hato}(1996)}]{Hato1996Jan}%
  \BibitemOpen
  \bibfield  {author} {\bibinfo {author} {\bibfnamefont {M.}~\bibnamefont
  {Hato}},\ }\href {\doibase 10.1021/jp961927h} {\bibfield  {journal} {\bibinfo
   {journal} {J. Phys. Chem.}\ }\textbf {\bibinfo {volume} {100}},\ \bibinfo
  {pages} {18530} (\bibinfo {year} {1996})}\BibitemShut {NoStop}%
\bibitem [{\citenamefont {Meyer}, \citenamefont {Rosenberg},\ and\
  \citenamefont {Israelachvili}(2006)}]{Meyer2006}%
  \BibitemOpen
  \bibfield  {author} {\bibinfo {author} {\bibfnamefont {E.~E.}\ \bibnamefont
  {Meyer}}, \bibinfo {author} {\bibfnamefont {K.~J.}\ \bibnamefont
  {Rosenberg}}, \ and\ \bibinfo {author} {\bibfnamefont {J.}~\bibnamefont
  {Israelachvili}},\ }\href {\doibase 10.1073/pnas.0606422103} {\bibfield
  {journal} {\bibinfo  {journal} {Proc. Natl. Acad. Sci. U.S.A.}\ }\textbf
  {\bibinfo {volume} {103}},\ \bibinfo {pages} {15739} (\bibinfo {year}
  {2006})}\BibitemShut {NoStop}%
\bibitem [{\citenamefont {Dishon}, \citenamefont {Zohar},\ and\ \citenamefont
  {Sivan}(2009)}]{Dishon2009}%
  \BibitemOpen
  \bibfield  {author} {\bibinfo {author} {\bibfnamefont {M.}~\bibnamefont
  {Dishon}}, \bibinfo {author} {\bibfnamefont {O.}~\bibnamefont {Zohar}}, \
  and\ \bibinfo {author} {\bibfnamefont {U.}~\bibnamefont {Sivan}},\ }\href
  {\doibase 10.1021/la803022b} {\bibfield  {journal} {\bibinfo  {journal}
  {Langmuir}\ }\textbf {\bibinfo {volume} {25}},\ \bibinfo {pages} {2831}
  (\bibinfo {year} {2009})}\BibitemShut {NoStop}%
\bibitem [{\citenamefont {Mastropietro}\ and\ \citenamefont
  {Ducker}(2012)}]{Mastropietro2012Mar}%
  \BibitemOpen
  \bibfield  {author} {\bibinfo {author} {\bibfnamefont {D.~J.}\ \bibnamefont
  {Mastropietro}}\ and\ \bibinfo {author} {\bibfnamefont {W.~A.}\ \bibnamefont
  {Ducker}},\ }\href {\doibase 10.1103/PhysRevLett.108.106101} {\bibfield
  {journal} {\bibinfo  {journal} {Phys. Rev. Lett.}\ }\textbf {\bibinfo
  {volume} {108}},\ \bibinfo {pages} {106101} (\bibinfo {year}
  {2012})}\BibitemShut {NoStop}%
\bibitem [{\citenamefont {Azadi}, \citenamefont {Nguyen},\ and\ \citenamefont
  {Yakubov}(2015)}]{Azadi2015Feb}%
  \BibitemOpen
  \bibfield  {author} {\bibinfo {author} {\bibfnamefont {M.}~\bibnamefont
  {Azadi}}, \bibinfo {author} {\bibfnamefont {A.~V.}\ \bibnamefont {Nguyen}}, \
  and\ \bibinfo {author} {\bibfnamefont {G.~E.}\ \bibnamefont {Yakubov}},\
  }\href {\doibase 10.1021/la504001z} {\bibfield  {journal} {\bibinfo
  {journal} {Langmuir}\ }\textbf {\bibinfo {volume} {31}},\ \bibinfo {pages}
  {1941} (\bibinfo {year} {2015})}\BibitemShut {NoStop}%
\bibitem [{\citenamefont {Schlesinger}\ and\ \citenamefont
  {Sivan}(2017)}]{Schlesinger2017Mar}%
  \BibitemOpen
  \bibfield  {author} {\bibinfo {author} {\bibfnamefont {I.}~\bibnamefont
  {Schlesinger}}\ and\ \bibinfo {author} {\bibfnamefont {U.}~\bibnamefont
  {Sivan}},\ }\href {\doibase 10.1021/acs.langmuir.6b03574} {\bibfield
  {journal} {\bibinfo  {journal} {Langmuir}\ }\textbf {\bibinfo {volume}
  {33}},\ \bibinfo {pages} {2485} (\bibinfo {year} {2017})}\BibitemShut
  {NoStop}%
\bibitem [{\citenamefont {Ishida}\ \emph {et~al.}(2018)\citenamefont {Ishida},
  \citenamefont {Matsuo}, \citenamefont {Imamura},\ and\ \citenamefont
  {Craig}}]{Ishida2018Feb}%
  \BibitemOpen
  \bibfield  {author} {\bibinfo {author} {\bibfnamefont {N.}~\bibnamefont
  {Ishida}}, \bibinfo {author} {\bibfnamefont {K.}~\bibnamefont {Matsuo}},
  \bibinfo {author} {\bibfnamefont {K.}~\bibnamefont {Imamura}}, \ and\
  \bibinfo {author} {\bibfnamefont {V.~S.~J.}\ \bibnamefont {Craig}},\
  }\href@noop {} {\bibfield  {journal} {\bibinfo  {journal} {Langmuir}\ }
  (\bibinfo {year} {2018})}\BibitemShut {NoStop}%
\bibitem [{\citenamefont {Marra}\ and\ \citenamefont
  {Israelachvili}(1985)}]{Marra1985Aug}%
  \BibitemOpen
  \bibfield  {author} {\bibinfo {author} {\bibfnamefont {J.}~\bibnamefont
  {Marra}}\ and\ \bibinfo {author} {\bibfnamefont {J.}~\bibnamefont
  {Israelachvili}},\ }\href {\doibase 10.1021/bi00338a020} {\bibfield
  {journal} {\bibinfo  {journal} {Biochemistry}\ }\textbf {\bibinfo {volume}
  {24}},\ \bibinfo {pages} {4608} (\bibinfo {year} {1985})}\BibitemShut
  {NoStop}%
\bibitem [{\citenamefont {Israelachvili}\ and\ \citenamefont
  {Pashley}(1983)}]{Israelachvili1983Nov}%
  \BibitemOpen
  \bibfield  {author} {\bibinfo {author} {\bibfnamefont {J.~N.}\ \bibnamefont
  {Israelachvili}}\ and\ \bibinfo {author} {\bibfnamefont {R.~M.}\ \bibnamefont
  {Pashley}},\ }\href {\doibase 10.1038/306249a0} {\bibfield  {journal}
  {\bibinfo  {journal} {Nature}\ }\textbf {\bibinfo {volume} {306}},\ \bibinfo
  {pages} {249} (\bibinfo {year} {1983})}\BibitemShut {NoStop}%
\bibitem [{\citenamefont {Chen}\ \emph {et~al.}(2016)\citenamefont {Chen},
  \citenamefont {Cox}, \citenamefont {Ow}, \citenamefont {Shi},\ and\
  \citenamefont {Panagiotopoulos}}]{Chen2016Jun}%
  \BibitemOpen
  \bibfield  {author} {\bibinfo {author} {\bibfnamefont {H.}~\bibnamefont
  {Chen}}, \bibinfo {author} {\bibfnamefont {J.~R.}\ \bibnamefont {Cox}},
  \bibinfo {author} {\bibfnamefont {H.}~\bibnamefont {Ow}}, \bibinfo {author}
  {\bibfnamefont {R.}~\bibnamefont {Shi}}, \ and\ \bibinfo {author}
  {\bibfnamefont {A.~Z.}\ \bibnamefont {Panagiotopoulos}},\ }\href {\doibase
  10.1038/srep28553} {\bibfield  {journal} {\bibinfo  {journal} {Sci. Rep.}\
  }\textbf {\bibinfo {volume} {6}},\ \bibinfo {pages} {28553} (\bibinfo {year}
  {2016})}\BibitemShut {NoStop}%
\bibitem [{\citenamefont {Vaikuntanathan}\ \emph {et~al.}(2016)\citenamefont
  {Vaikuntanathan}, \citenamefont {Rotskoff}, \citenamefont {Hudson},\ and\
  \citenamefont {Geissler}}]{Vaikuntanathan2016Apr}%
  \BibitemOpen
  \bibfield  {author} {\bibinfo {author} {\bibfnamefont {S.}~\bibnamefont
  {Vaikuntanathan}}, \bibinfo {author} {\bibfnamefont {G.}~\bibnamefont
  {Rotskoff}}, \bibinfo {author} {\bibfnamefont {A.}~\bibnamefont {Hudson}}, \
  and\ \bibinfo {author} {\bibfnamefont {P.~L.}\ \bibnamefont {Geissler}},\
  }\href {\doibase 10.1073/pnas.1513659113} {\bibfield  {journal} {\bibinfo
  {journal} {Proc. Natl. Acad. Sci. U.S.A.}\ }\textbf {\bibinfo {volume}
  {113}},\ \bibinfo {pages} {E2224} (\bibinfo {year} {2016})}\BibitemShut
  {NoStop}%
\bibitem [{\citenamefont {Bolhuis}\ and\ \citenamefont
  {Chandler}(2000)}]{Bolhuis2000Nov}%
  \BibitemOpen
  \bibfield  {author} {\bibinfo {author} {\bibfnamefont {P.~G.}\ \bibnamefont
  {Bolhuis}}\ and\ \bibinfo {author} {\bibfnamefont {D.}~\bibnamefont
  {Chandler}},\ }\href {\doibase 10.1063/1.1315997} {\bibfield  {journal}
  {\bibinfo  {journal} {J. Chem. Phys.}\ }\textbf {\bibinfo {volume} {113}},\
  \bibinfo {pages} {8154} (\bibinfo {year} {2000})}\BibitemShut {NoStop}%
\bibitem [{\citenamefont {Altabet}, \citenamefont {Haji-Akbari},\ and\
  \citenamefont {Debenedetti}(2017)}]{Altabet2017Mar}%
  \BibitemOpen
  \bibfield  {author} {\bibinfo {author} {\bibfnamefont {Y.~E.}\ \bibnamefont
  {Altabet}}, \bibinfo {author} {\bibfnamefont {A.}~\bibnamefont
  {Haji-Akbari}}, \ and\ \bibinfo {author} {\bibfnamefont {P.~G.}\ \bibnamefont
  {Debenedetti}},\ }\href {\doibase 10.1073/pnas.1620335114} {\bibfield
  {journal} {\bibinfo  {journal} {Proc. Natl. Acad. Sci. U.S.A.}\ }\textbf
  {\bibinfo {volume} {114}},\ \bibinfo {pages} {E2548} (\bibinfo {year}
  {2017})}\BibitemShut {NoStop}%
\bibitem [{\citenamefont {Remsing}\ \emph {et~al.}(2015)\citenamefont
  {Remsing}, \citenamefont {Xi}, \citenamefont {Vembanur}, \citenamefont
  {Sharma}, \citenamefont {Debenedetti}, \citenamefont {Garde},\ and\
  \citenamefont {Patel}}]{Remsing2015Jul}%
  \BibitemOpen
  \bibfield  {author} {\bibinfo {author} {\bibfnamefont {R.~C.}\ \bibnamefont
  {Remsing}}, \bibinfo {author} {\bibfnamefont {E.}~\bibnamefont {Xi}},
  \bibinfo {author} {\bibfnamefont {S.}~\bibnamefont {Vembanur}}, \bibinfo
  {author} {\bibfnamefont {S.}~\bibnamefont {Sharma}}, \bibinfo {author}
  {\bibfnamefont {P.~G.}\ \bibnamefont {Debenedetti}}, \bibinfo {author}
  {\bibfnamefont {S.}~\bibnamefont {Garde}}, \ and\ \bibinfo {author}
  {\bibfnamefont {A.~J.}\ \bibnamefont {Patel}},\ }\href {\doibase
  10.1073/pnas.1503302112} {\bibfield  {journal} {\bibinfo  {journal} {Proc.
  Natl. Acad. Sci. U.S.A.}\ }\textbf {\bibinfo {volume} {112}},\ \bibinfo
  {pages} {8181} (\bibinfo {year} {2015})}\BibitemShut {NoStop}%
\bibitem [{\citenamefont {B{\ifmmode \acute{e} \else \'{e}\fi}rard},
  \citenamefont {Attard},\ and\ \citenamefont {Patey}(1993)}]{Berard1993May}%
  \BibitemOpen
  \bibfield  {author} {\bibinfo {author} {\bibfnamefont {D.~R.}\ \bibnamefont
  {B{\ifmmode \acute{e} \else \'{e}\fi}rard}}, \bibinfo {author} {\bibfnamefont
  {P.}~\bibnamefont {Attard}}, \ and\ \bibinfo {author} {\bibfnamefont {G.~N.}\
  \bibnamefont {Patey}},\ }\href {\doibase 10.1063/1.464715} {\bibfield
  {journal} {\bibinfo  {journal} {J. Chem. Phys.}\ }\textbf {\bibinfo {volume}
  {98}},\ \bibinfo {pages} {7236} (\bibinfo {year} {1993})}\BibitemShut
  {NoStop}%
\bibitem [{\citenamefont {Evans}, \citenamefont {Stewart},\ and\ \citenamefont
  {Wilding}(2017)}]{Evans2017Jul}%
  \BibitemOpen
  \bibfield  {author} {\bibinfo {author} {\bibfnamefont {R.}~\bibnamefont
  {Evans}}, \bibinfo {author} {\bibfnamefont {M.~C.}\ \bibnamefont {Stewart}},
  \ and\ \bibinfo {author} {\bibfnamefont {N.~B.}\ \bibnamefont {Wilding}},\
  }\href {\doibase 10.1063/1.4993515} {\bibfield  {journal} {\bibinfo
  {journal} {J. Chem. Phys.}\ }\textbf {\bibinfo {volume} {147}},\ \bibinfo
  {pages} {044701} (\bibinfo {year} {2017})}\BibitemShut {NoStop}%
\bibitem [{\citenamefont {Qin}\ and\ \citenamefont
  {Fichthorn}(2003)}]{Qin2003Nov}%
  \BibitemOpen
  \bibfield  {author} {\bibinfo {author} {\bibfnamefont {Y.}~\bibnamefont
  {Qin}}\ and\ \bibinfo {author} {\bibfnamefont {K.~A.}\ \bibnamefont
  {Fichthorn}},\ }\href {\doibase 10.1063/1.1615493} {\bibfield  {journal}
  {\bibinfo  {journal} {J. Chem. Phys.}\ }\textbf {\bibinfo {volume} {119}},\
  \bibinfo {pages} {9745} (\bibinfo {year} {2003})}\BibitemShut {NoStop}%
\bibitem [{\citenamefont {Jabes}, \citenamefont {Bratko},\ and\ \citenamefont
  {Luzar}(2016)}]{Jabes2016Aug}%
  \BibitemOpen
  \bibfield  {author} {\bibinfo {author} {\bibfnamefont {B.~S.}\ \bibnamefont
  {Jabes}}, \bibinfo {author} {\bibfnamefont {D.}~\bibnamefont {Bratko}}, \
  and\ \bibinfo {author} {\bibfnamefont {A.}~\bibnamefont {Luzar}},\ }\href
  {\doibase 10.1021/acs.jpclett.6b01442} {\bibfield  {journal} {\bibinfo
  {journal} {J. Phys. Chem. Lett.}\ }\textbf {\bibinfo {volume} {7}},\ \bibinfo
  {pages} {3158} (\bibinfo {year} {2016})}\BibitemShut {NoStop}%
\bibitem [{\citenamefont {Qin}\ and\ \citenamefont
  {Fichthorn}(2006)}]{Qin2006Feb}%
  \BibitemOpen
  \bibfield  {author} {\bibinfo {author} {\bibfnamefont {Y.}~\bibnamefont
  {Qin}}\ and\ \bibinfo {author} {\bibfnamefont {K.~A.}\ \bibnamefont
  {Fichthorn}},\ }\href {\doibase 10.1103/PhysRevE.73.020401} {\bibfield
  {journal} {\bibinfo  {journal} {Phys. Rev. E}\ }\textbf {\bibinfo {volume}
  {73}},\ \bibinfo {pages} {020401} (\bibinfo {year} {2006})}\BibitemShut
  {NoStop}%
\bibitem [{\citenamefont {Lum}, \citenamefont {Chandler},\ and\ \citenamefont
  {Weeks}(1999)}]{Lum1999Jun}%
  \BibitemOpen
  \bibfield  {author} {\bibinfo {author} {\bibfnamefont {K.}~\bibnamefont
  {Lum}}, \bibinfo {author} {\bibfnamefont {D.}~\bibnamefont {Chandler}}, \
  and\ \bibinfo {author} {\bibfnamefont {J.~D.}\ \bibnamefont {Weeks}},\ }\href
  {\doibase 10.1021/jp984327m} {\bibfield  {journal} {\bibinfo  {journal} {J.
  Phys. Chem. B}\ }\textbf {\bibinfo {volume} {103}},\ \bibinfo {pages} {4570}
  (\bibinfo {year} {1999})}\BibitemShut {NoStop}%
\bibitem [{\citenamefont {Varilly}, \citenamefont {Patel},\ and\ \citenamefont
  {Chandler}(2011)}]{Varilly2011Feb}%
  \BibitemOpen
  \bibfield  {author} {\bibinfo {author} {\bibfnamefont {P.}~\bibnamefont
  {Varilly}}, \bibinfo {author} {\bibfnamefont {A.~J.}\ \bibnamefont {Patel}},
  \ and\ \bibinfo {author} {\bibfnamefont {D.}~\bibnamefont {Chandler}},\
  }\href {\doibase 10.1063/1.3532939} {\bibfield  {journal} {\bibinfo
  {journal} {J. Chem. Phys.}\ }\textbf {\bibinfo {volume} {134}},\ \bibinfo
  {pages} {074109} (\bibinfo {year} {2011})}\BibitemShut {NoStop}%
\bibitem [{\citenamefont {Remsing}\ and\ \citenamefont
  {Weeks}(2013)}]{Remsing2013Dec}%
  \BibitemOpen
  \bibfield  {author} {\bibinfo {author} {\bibfnamefont {R.~C.}\ \bibnamefont
  {Remsing}}\ and\ \bibinfo {author} {\bibfnamefont {J.~D.}\ \bibnamefont
  {Weeks}},\ }\href {\doibase 10.1021/jp4053067} {\bibfield  {journal}
  {\bibinfo  {journal} {J. Phys. Chem. B}\ }\textbf {\bibinfo {volume} {117}},\
  \bibinfo {pages} {15479} (\bibinfo {year} {2013})}\BibitemShut {NoStop}%
\bibitem [{\citenamefont {Asakura}\ and\ \citenamefont
  {Oosawa}(1954)}]{asakura1954}%
  \BibitemOpen
  \bibfield  {author} {\bibinfo {author} {\bibfnamefont {S.}~\bibnamefont
  {Asakura}}\ and\ \bibinfo {author} {\bibfnamefont {F.}~\bibnamefont
  {Oosawa}},\ }\href {\doibase 10.1063/1.1740347} {\bibfield  {journal}
  {\bibinfo  {journal} {The Journal of Chemical Physics}\ }\textbf {\bibinfo
  {volume} {22}},\ \bibinfo {pages} {1255} (\bibinfo {year}
  {1954})}\BibitemShut {NoStop}%
\bibitem [{\citenamefont {Dickman}, \citenamefont {Attard},\ and\ \citenamefont
  {Simonian}(1997)}]{dickman1997}%
  \BibitemOpen
  \bibfield  {author} {\bibinfo {author} {\bibfnamefont {R.}~\bibnamefont
  {Dickman}}, \bibinfo {author} {\bibfnamefont {P.}~\bibnamefont {Attard}}, \
  and\ \bibinfo {author} {\bibfnamefont {V.}~\bibnamefont {Simonian}},\ }\href
  {\doibase 10.1063/1.474367} {\bibfield  {journal} {\bibinfo  {journal} {The
  Journal of chemical physics}\ }\textbf {\bibinfo {volume} {107}},\ \bibinfo
  {pages} {205} (\bibinfo {year} {1997})}\BibitemShut {NoStop}%
\bibitem [{\citenamefont {Roth}, \citenamefont {G{\"o}tzelmann},\ and\
  \citenamefont {Dietrich}(1999)}]{roth1999}%
  \BibitemOpen
  \bibfield  {author} {\bibinfo {author} {\bibfnamefont {R.}~\bibnamefont
  {Roth}}, \bibinfo {author} {\bibfnamefont {B.}~\bibnamefont
  {G{\"o}tzelmann}}, \ and\ \bibinfo {author} {\bibfnamefont {S.}~\bibnamefont
  {Dietrich}},\ }\href {\doibase 10.1103/PhysRevLett.83.448} {\bibfield
  {journal} {\bibinfo  {journal} {Physical review letters}\ }\textbf {\bibinfo
  {volume} {83}},\ \bibinfo {pages} {448} (\bibinfo {year} {1999})}\BibitemShut
  {NoStop}%
\bibitem [{\citenamefont {Xiao}, \citenamefont {Guo},\ and\ \citenamefont
  {Li}(2005)}]{xiao2005}%
  \BibitemOpen
  \bibfield  {author} {\bibinfo {author} {\bibfnamefont {C.}~\bibnamefont
  {Xiao}}, \bibinfo {author} {\bibfnamefont {J.}~\bibnamefont {Guo}}, \ and\
  \bibinfo {author} {\bibfnamefont {C.}~\bibnamefont {Li}},\ }\href {\doibase
  10.1209/epl/i2005-10397-2} {\bibfield  {journal} {\bibinfo  {journal} {EPL
  (Europhysics Letters)}\ }\textbf {\bibinfo {volume} {73}},\ \bibinfo {pages}
  {443} (\bibinfo {year} {2005})}\BibitemShut {NoStop}%
\bibitem [{\citenamefont {Nyg{\aa}rd}\ \emph {et~al.}(2016)\citenamefont
  {Nyg{\aa}rd}, \citenamefont {Sarman}, \citenamefont {Hyltegren},
  \citenamefont {Chodankar}, \citenamefont {Perret}, \citenamefont
  {Buitenhuis}, \citenamefont {van~der Veen},\ and\ \citenamefont
  {Kjellander}}]{nygaard2016a}%
  \BibitemOpen
  \bibfield  {author} {\bibinfo {author} {\bibfnamefont {K.}~\bibnamefont
  {Nyg{\aa}rd}}, \bibinfo {author} {\bibfnamefont {S.}~\bibnamefont {Sarman}},
  \bibinfo {author} {\bibfnamefont {K.}~\bibnamefont {Hyltegren}}, \bibinfo
  {author} {\bibfnamefont {S.}~\bibnamefont {Chodankar}}, \bibinfo {author}
  {\bibfnamefont {E.}~\bibnamefont {Perret}}, \bibinfo {author} {\bibfnamefont
  {J.}~\bibnamefont {Buitenhuis}}, \bibinfo {author} {\bibfnamefont {J.~F.}\
  \bibnamefont {van~der Veen}}, \ and\ \bibinfo {author} {\bibfnamefont
  {R.}~\bibnamefont {Kjellander}},\ }\href {\doibase 10.1103/PhysRevX.6.011014}
  {\bibfield  {journal} {\bibinfo  {journal} {Physical Review X}\ }\textbf
  {\bibinfo {volume} {6}},\ \bibinfo {pages} {011014} (\bibinfo {year}
  {2016})}\BibitemShut {NoStop}%
\bibitem [{\citenamefont {Nyg{\aa}rd}(2016)}]{nygaard2016b}%
  \BibitemOpen
  \bibfield  {author} {\bibinfo {author} {\bibfnamefont {K.}~\bibnamefont
  {Nyg{\aa}rd}},\ }\href {\doibase 10.1016/j.cocis.2016.02.005} {\bibfield
  {journal} {\bibinfo  {journal} {Current Opinion in Colloid \& Interface
  Science}\ }\textbf {\bibinfo {volume} {22}},\ \bibinfo {pages} {30} (\bibinfo
  {year} {2016})}\BibitemShut {NoStop}%
\bibitem [{\citenamefont {Mishima}\ \emph
  {et~al.}(2013{\natexlab{a}})\citenamefont {Mishima}, \citenamefont {Oshima},
  \citenamefont {Yasuda}, \citenamefont {Amano},\ and\ \citenamefont
  {Kinoshita}}]{mishima2013a}%
  \BibitemOpen
  \bibfield  {author} {\bibinfo {author} {\bibfnamefont {H.}~\bibnamefont
  {Mishima}}, \bibinfo {author} {\bibfnamefont {H.}~\bibnamefont {Oshima}},
  \bibinfo {author} {\bibfnamefont {S.}~\bibnamefont {Yasuda}}, \bibinfo
  {author} {\bibfnamefont {K.-i.}\ \bibnamefont {Amano}}, \ and\ \bibinfo
  {author} {\bibfnamefont {M.}~\bibnamefont {Kinoshita}},\ }\href@noop {}
  {\bibfield  {journal} {\bibinfo  {journal} {The Journal of chemical physics}\
  }\textbf {\bibinfo {volume} {139}},\ \bibinfo {pages} {11B618\_1} (\bibinfo
  {year} {2013}{\natexlab{a}})}\BibitemShut {NoStop}%
\bibitem [{\citenamefont {Mishima}\ \emph
  {et~al.}(2013{\natexlab{b}})\citenamefont {Mishima}, \citenamefont {Oshima},
  \citenamefont {Yasuda}, \citenamefont {Amano},\ and\ \citenamefont
  {Kinoshita}}]{mishima2013b}%
  \BibitemOpen
  \bibfield  {author} {\bibinfo {author} {\bibfnamefont {H.}~\bibnamefont
  {Mishima}}, \bibinfo {author} {\bibfnamefont {H.}~\bibnamefont {Oshima}},
  \bibinfo {author} {\bibfnamefont {S.}~\bibnamefont {Yasuda}}, \bibinfo
  {author} {\bibfnamefont {K.-i.}\ \bibnamefont {Amano}}, \ and\ \bibinfo
  {author} {\bibfnamefont {M.}~\bibnamefont {Kinoshita}},\ }\href@noop {}
  {\bibfield  {journal} {\bibinfo  {journal} {Chemical Physics Letters}\
  }\textbf {\bibinfo {volume} {561}},\ \bibinfo {pages} {159} (\bibinfo {year}
  {2013}{\natexlab{b}})}\BibitemShut {NoStop}%
\bibitem [{\citenamefont {Hara}\ \emph {et~al.}(2016)\citenamefont {Hara},
  \citenamefont {Amano}, \citenamefont {Kinoshita},\ and\ \citenamefont
  {Yoshimori}}]{hara2016}%
  \BibitemOpen
  \bibfield  {author} {\bibinfo {author} {\bibfnamefont {R.}~\bibnamefont
  {Hara}}, \bibinfo {author} {\bibfnamefont {K.-i.}\ \bibnamefont {Amano}},
  \bibinfo {author} {\bibfnamefont {M.}~\bibnamefont {Kinoshita}}, \ and\
  \bibinfo {author} {\bibfnamefont {A.}~\bibnamefont {Yoshimori}},\ }\href
  {\doibase 10.1063/1.4943394} {\bibfield  {journal} {\bibinfo  {journal} {The
  Journal of chemical physics}\ }\textbf {\bibinfo {volume} {144}},\ \bibinfo
  {pages} {105103} (\bibinfo {year} {2016})}\BibitemShut {NoStop}%
\bibitem [{\citenamefont {Stewart}\ and\ \citenamefont
  {Evans}(2014)}]{stewart2014}%
  \BibitemOpen
  \bibfield  {author} {\bibinfo {author} {\bibfnamefont {M.~C.}\ \bibnamefont
  {Stewart}}\ and\ \bibinfo {author} {\bibfnamefont {R.}~\bibnamefont
  {Evans}},\ }\href {\doibase 10.1063/1.4869868} {\bibfield  {journal}
  {\bibinfo  {journal} {The Journal of chemical physics}\ }\textbf {\bibinfo
  {volume} {140}},\ \bibinfo {pages} {134704} (\bibinfo {year}
  {2014})}\BibitemShut {NoStop}%
\bibitem [{\citenamefont {Macio{\l}ek}, \citenamefont {Drzewi{\'n}ski},\ and\
  \citenamefont {Bryk}(2004)}]{maciolek2004}%
  \BibitemOpen
  \bibfield  {author} {\bibinfo {author} {\bibfnamefont {A.}~\bibnamefont
  {Macio{\l}ek}}, \bibinfo {author} {\bibfnamefont {A.}~\bibnamefont
  {Drzewi{\'n}ski}}, \ and\ \bibinfo {author} {\bibfnamefont {P.}~\bibnamefont
  {Bryk}},\ }\href {\doibase 10.1063/1.1635807} {\bibfield  {journal} {\bibinfo
   {journal} {The Journal of chemical physics}\ }\textbf {\bibinfo {volume}
  {120}},\ \bibinfo {pages} {1921} (\bibinfo {year} {2004})}\BibitemShut
  {NoStop}%
\bibitem [{\citenamefont {Lutsko}(2010)}]{Lutsko2010Jan}%
  \BibitemOpen
  \bibfield  {author} {\bibinfo {author} {\bibfnamefont {J.~F.}\ \bibnamefont
  {Lutsko}},\ }\href {\doibase 10.1002/9780470564318.ch1} {\enquote {\bibinfo
  {title} {{Recent Developments in Classical Density Functional Theory}},}\ }
  (\bibinfo {year} {2010}),\ \bibinfo {note} {[Online; accessed 16. Apr.
  2018]}\BibitemShut {NoStop}%
\bibitem [{\citenamefont {Jeanmairet}\ \emph {et~al.}(2013)\citenamefont
  {Jeanmairet}, \citenamefont {Levesque}, \citenamefont {Vuilleumier},\ and\
  \citenamefont {Borgis}}]{Jeanmairet2013Feb}%
  \BibitemOpen
  \bibfield  {author} {\bibinfo {author} {\bibfnamefont {G.}~\bibnamefont
  {Jeanmairet}}, \bibinfo {author} {\bibfnamefont {M.}~\bibnamefont
  {Levesque}}, \bibinfo {author} {\bibfnamefont {R.}~\bibnamefont
  {Vuilleumier}}, \ and\ \bibinfo {author} {\bibfnamefont {D.}~\bibnamefont
  {Borgis}},\ }\href {\doibase 10.1021/jz301956b} {\bibfield  {journal}
  {\bibinfo  {journal} {J. Phys. Chem. Lett.}\ }\textbf {\bibinfo {volume}
  {4}},\ \bibinfo {pages} {619} (\bibinfo {year} {2013})}\BibitemShut {NoStop}%
\bibitem [{\citenamefont {Jeanmairet}, \citenamefont {Levesque},\ and\
  \citenamefont {Borgis}(2013)}]{Jeanmairet2013Oct}%
  \BibitemOpen
  \bibfield  {author} {\bibinfo {author} {\bibfnamefont {G.}~\bibnamefont
  {Jeanmairet}}, \bibinfo {author} {\bibfnamefont {M.}~\bibnamefont
  {Levesque}}, \ and\ \bibinfo {author} {\bibfnamefont {D.}~\bibnamefont
  {Borgis}},\ }\href {\doibase 10.1063/1.4824737} {\bibfield  {journal}
  {\bibinfo  {journal} {J. Chem. Phys.}\ }\textbf {\bibinfo {volume} {139}},\
  \bibinfo {pages} {154101} (\bibinfo {year} {2013})}\BibitemShut {NoStop}%
\bibitem [{\citenamefont {Hughes}, \citenamefont {Krebs},\ and\ \citenamefont
  {Roundy}(2013)}]{Hughes2013Jan}%
  \BibitemOpen
  \bibfield  {author} {\bibinfo {author} {\bibfnamefont {J.}~\bibnamefont
  {Hughes}}, \bibinfo {author} {\bibfnamefont {E.~J.}\ \bibnamefont {Krebs}}, \
  and\ \bibinfo {author} {\bibfnamefont {D.}~\bibnamefont {Roundy}},\ }\href
  {\doibase 10.1063/1.4774155} {\bibfield  {journal} {\bibinfo  {journal} {J.
  Chem. Phys.}\ }\textbf {\bibinfo {volume} {138}},\ \bibinfo {pages} {024509}
  (\bibinfo {year} {2013})}\BibitemShut {NoStop}%
\bibitem [{\citenamefont {Ashbaugh}(2013)}]{Ashbaugh2013Aug}%
  \BibitemOpen
  \bibfield  {author} {\bibinfo {author} {\bibfnamefont {H.~S.}\ \bibnamefont
  {Ashbaugh}},\ }\href {\doibase 10.1063/1.4817661} {\bibfield  {journal}
  {\bibinfo  {journal} {J. Chem. Phys.}\ }\textbf {\bibinfo {volume} {139}},\
  \bibinfo {pages} {064702} (\bibinfo {year} {2013})}\BibitemShut {NoStop}%
\bibitem [{\citenamefont {Evans}\ and\ \citenamefont
  {Stewart}(2015)}]{Evans2015Apr}%
  \BibitemOpen
  \bibfield  {author} {\bibinfo {author} {\bibfnamefont {R.}~\bibnamefont
  {Evans}}\ and\ \bibinfo {author} {\bibfnamefont {M.~C.}\ \bibnamefont
  {Stewart}},\ }\href {\doibase 10.1088/0953-8984/27/19/194111} {\bibfield
  {journal} {\bibinfo  {journal} {J. Phys.: Condens. Matter}\ }\textbf
  {\bibinfo {volume} {27}},\ \bibinfo {pages} {194111} (\bibinfo {year}
  {2015})}\BibitemShut {NoStop}%
\bibitem [{\citenamefont {Evans}\ and\ \citenamefont
  {Wilding}(2015)}]{Evans2015Jul}%
  \BibitemOpen
  \bibfield  {author} {\bibinfo {author} {\bibfnamefont {R.}~\bibnamefont
  {Evans}}\ and\ \bibinfo {author} {\bibfnamefont {N.~B.}\ \bibnamefont
  {Wilding}},\ }\href {\doibase 10.1103/PhysRevLett.115.016103} {\bibfield
  {journal} {\bibinfo  {journal} {Phys. Rev. Lett.}\ }\textbf {\bibinfo
  {volume} {115}},\ \bibinfo {pages} {016103} (\bibinfo {year}
  {2015})}\BibitemShut {NoStop}%
\bibitem [{\citenamefont {Ashcroft}\ and\ \citenamefont
  {Mermin}(1976)}]{Ashcroft}%
  \BibitemOpen
  \bibfield  {author} {\bibinfo {author} {\bibfnamefont {N.}~\bibnamefont
  {Ashcroft}}\ and\ \bibinfo {author} {\bibfnamefont {N.}~\bibnamefont
  {Mermin}},\ }\href@noop {} {\emph {\bibinfo {title} {{Solid State
  Physics}}}}\ (\bibinfo  {publisher} {Saunders College},\ \bibinfo {address}
  {Philadelphia},\ \bibinfo {year} {1976})\BibitemShut {NoStop}%
\bibitem [{\citenamefont {Berendsen}, \citenamefont {Grigera},\ and\
  \citenamefont {Straatsma}(1987)}]{Berendsen1987Nov}%
  \BibitemOpen
  \bibfield  {author} {\bibinfo {author} {\bibfnamefont {H.~J.~C.}\
  \bibnamefont {Berendsen}}, \bibinfo {author} {\bibfnamefont {J.~R.}\
  \bibnamefont {Grigera}}, \ and\ \bibinfo {author} {\bibfnamefont {T.~P.}\
  \bibnamefont {Straatsma}},\ }\href {\doibase 10.1021/j100308a038} {\bibfield
  {journal} {\bibinfo  {journal} {J. Phys. Chem.}\ }\textbf {\bibinfo {volume}
  {91}},\ \bibinfo {pages} {6269} (\bibinfo {year} {1987})}\BibitemShut
  {NoStop}%
\bibitem [{\citenamefont {Plimpton}(1995)}]{plimpton1995}%
  \BibitemOpen
  \bibfield  {author} {\bibinfo {author} {\bibfnamefont {S.}~\bibnamefont
  {Plimpton}},\ }\href {http://lammps.sandia.gov/index.html} {\bibfield
  {journal} {\bibinfo  {journal} {J. Comput. Phys.}\ }\textbf {\bibinfo
  {volume} {117}},\ \bibinfo {pages} {1} (\bibinfo {year} {1995})}\BibitemShut
  {NoStop}%
\bibitem [{\citenamefont {Roth}\ \emph {et~al.}(2002)\citenamefont {Roth},
  \citenamefont {Evans}, \citenamefont {Lang},\ and\ \citenamefont
  {Kahl}}]{Roth2002Nov}%
  \BibitemOpen
  \bibfield  {author} {\bibinfo {author} {\bibfnamefont {R.}~\bibnamefont
  {Roth}}, \bibinfo {author} {\bibfnamefont {R.}~\bibnamefont {Evans}},
  \bibinfo {author} {\bibfnamefont {A.}~\bibnamefont {Lang}}, \ and\ \bibinfo
  {author} {\bibfnamefont {G.}~\bibnamefont {Kahl}},\ }\href {\doibase
  10.1088/0953-8984/14/46/313} {\bibfield  {journal} {\bibinfo  {journal} {J.
  Phys.: Condens. Matter}\ }\textbf {\bibinfo {volume} {14}},\ \bibinfo {pages}
  {12063} (\bibinfo {year} {2002})}\BibitemShut {NoStop}%
\bibitem [{\citenamefont {Lutsko}\ and\ \citenamefont
  {Lam}(2018)}]{Lutsko2018May}%
  \BibitemOpen
  \bibfield  {author} {\bibinfo {author} {\bibfnamefont {J.~F.}\ \bibnamefont
  {Lutsko}}\ and\ \bibinfo {author} {\bibfnamefont {J.}~\bibnamefont {Lam}},\
  }\href@noop {} {\bibfield  {journal} {\bibinfo  {journal} {arXiv}\ }
  (\bibinfo {year} {2018})},\ \Eprint {http://arxiv.org/abs/1805.05673}
  {1805.05673} \BibitemShut {NoStop}%
\end{thebibliography}%

\end{document}